\documentclass[a4paper,scriptaddress,twocolumn, prd,showkeys,showpacs]{revtex4-1}

	\usepackage{amsmath}
	\usepackage{makeidx}
	\usepackage{amsfonts}

\makeindex

\begin{document}

\title{Considerations on Gravity as an Entropic Force and Entangled States}

\author{Everton M. C. Abreu$^{a,b}$}
\email{evertonabreu@ufrrj.br}
\author{Jorge Ananias Neto$^b$}
\email{jorge@fisica.ufjf.br}

\affiliation{$^a$Grupo de F\' isica Te\'orica e Matem\'atica F\' isica,  Departamento de  F\'{\i}sica, 
Universidade Federal Rural do Rio de Janeiro,
BR 465-07, 23890-971,  Serop\'edica, RJ, Brazil.\\
$^b$Departamento de F\'isica, ICE, Universidade Federal de Juiz de Fora,  36036-900, Juiz de Fora, MG, Brazil}


\begin{abstract}
\noindent Verlinde's ideas considered gravity as an emergent force originated from entropic concepts.  This hypothesis generated a huge number of papers through the last recent years concerning classical and quantum approaches about the issue.  In a recent paper Kobakhidze, using ultra-cold neutrons experiment, claimed that Verlinde's entropic gravity is not correct.  In this letter, by considering the Tsallis nonadditivity entropy concerning the holographic screen, where we assumed that the bits are entangled states, we showed that it is possible to confirm Verlinde's formalism.
\end{abstract}

\keywords{Entropic gravity, nonextensive statistics  theory}
\pacs{04.50.-h, 05.20.-y, 05.90.+m, 05.70.-a}
\maketitle


\vskip 1 cm

The results of Kobakhidze's approach (KA) \cite{Kob} disprove the entropic force idea suggested by Verlinde's (VE) formalism \cite{Ver}. The central question is that the entropy formula defined by VE formalism, in principle, leads to a quantum neutron mixed state whose main outcome disagrees with the results from the ultra-cold neutron experiment in the Earth gravitational field \cite{Nes}. However, Chaichian, Oksanen and Tureanu \cite{COT} have pointed out that the KA steps simplifies the VE formalism and assumes hypothesis that are not mentioned in VE theory. So, the aim of this letter is to show that when we consider that the entropy of the neutron-screen system no longer obeys the additive property, then the neutron can remain in a pure state and consequently, there is no more contradiction between the VE formalism and the ultra-cold neutron experiments. 

Our principal argument is to consider that the bits of the holographic screen form an entangled state, i.e., a quantum state that can not be decomposed by a simple product of uncorrelated states. In this situation, where there are strong interactions, several studies point out that the entropy defined by Tsallis \cite{Tsa, Tsa1,Tsa2, Tsa3}, denoted from now on as TT formalism, is more adequate than the standard Boltzmann-Gibbs (BG) \cite{GG} one. For example, considering a recent application, we can mention the generalized black hole formulation provided by Tsallis and Cirto \cite{tc}.  This study disclosed underlying features about an expanding Universe in entropic cosmology, as was investigated with detail by Komatsu and Kimura \cite{kk}.

Before we provide strong arguments in favor of VE formalism, let us review, in a short way, both the VE theory and the KA. The VE formalism is based on many works that link the gravitational theory to thermodynamics. Among them,  we can mention studies by Bekenstein \cite{Bek}, Hawking \cite{Haw}, Unruh \cite{Unr} and Jacobson \cite{Jac}. We must also mention that related ideas to VE formalism were also presented by Padmanabhan \cite{Pad}. 

The VE concept considers a test massive particle approaching the holographic screen. An holographic screen is a storage device for information which is constituted by bits (bit  is the smallest unit of information) and it also shows thermodynamics properties. If the distance between the particle and holographic screen is $\Delta x$, then the holographic screen experiences an entropy variation given by

\begin{eqnarray}
\label{env}
\Delta S= 2\pi k_B \frac{m c}{\hbar}\, \Delta x.
\end{eqnarray}
The entropic force applied on the test particle has the following form

\begin{eqnarray}
\label{enf}
F \Delta x= T \Delta S.
\end{eqnarray}
Also, using the Unruh temperature 

\begin{eqnarray}
\label{Unr}
k_B T = \frac{1}{2\pi} \frac{\hbar a}{c}
\end{eqnarray}
and Eqs. (\ref{env}) and (\ref{enf}), we can obtain that the entropic force reduces to

\begin{eqnarray}
\label{enf2}
F = T \, \frac{\Delta S}{\Delta x} = m \, a,
\end{eqnarray}
which is just the Newton second law. So, we can see that from the thermodynamical point of view we are able to reproduce using simple arguments one of the most important equations in physics. Besides, using the equipartition theorem and the holographic principle, it is possible to derive the Newton law of gravity \cite{Ver}.

\bigskip

In a different way, the KA initially considers that the holographic screen contains a large number of microscopic states denoted by $| i(z)>$ and it is described by a mixed state

\begin{eqnarray}
\label{ms}
\rho_S(z)=\sum_{i(z)} p_{i(z)} \left| i(z)><i(z) \right|,
\end{eqnarray}
where $p_{i(z)}$ are the weights and satisfy the relations

\begin{eqnarray}
0<p_{i(z)}\leq 1; \;\;\; \sum_{i(z)} p_{i(z)}=1.
\end{eqnarray}
It is important to mention here that a quantum mechanical treatment of the holographic screen is, nowadays, poorly studied and consequently, it is an open question \cite{COT}. Therefore, it is reasonable to assume that the bits quantum states, contrarily to what was considered by KA,  can be more complex than a simple product of states. These states can constitute a quantum entanglement.  We will see that, according to TT formalism,  the additive property of entropy can be lost. 

The standard Von Neumann entropy (which is additive for a physical system that presents no interaction between its constituents)
is written as

\begin{eqnarray}
\label{VNS}
S(z)= -\,k_B Tr [ \rho(z) \ln{\rho(z)} ].
\end{eqnarray}
Let a particle (for instance, a neutron) be also described by a density operator $\rho_{\cal N}$ in which Von Neumann entropy is given by
\begin{eqnarray}
\label{VNN}
S = -\,k_B Tr[ \rho_{\cal N}\,\ln{\rho_{\cal N}} ].
\end{eqnarray}
After taking into account some considerations, we have the following equation for the neutron-screen systems \cite{Kob}

\begin{eqnarray}
\label{K1}
S_{\cal N}(z + \Delta z)=S_{\cal S}(z+\Delta z)-S_{\cal S/N}(z),
\end{eqnarray}
where $ S_{\cal N}$ is the entropy of the neutron, $S_{\cal S}$ is the entropy of the holographic screen, $S_{\cal S/N}$ is the entropy of the holographic screen without the fragment $\cal N$, denoted the ``coarse-grained" screen, and $\Delta z$ is the distance between the neutron and the holographic screen. The neutron fragments in the holographic screen are bits which contain information about the neutron particle. For small $\Delta z$, the entropy $S_{\cal S}(z+\Delta z)$ can be expanded as 

\begin{eqnarray}
\label{ene}
S_{\cal S}(z+\Delta z)\approx S_{\cal S}(z)+\Delta S_{\cal S}.
\end{eqnarray}
KA assumes that the entropy of holographic screen is additive. Then we have

\begin{eqnarray}
\label{ena}
S_{\cal S/N}(z)=S_{\cal S}(z)-S_{\cal N}(z).
\end{eqnarray}
Substituting Eqs. (\ref{ene}) and (\ref{ena}) into Eq. (\ref{K1}), we obtain

\begin{eqnarray}
\label{enm}
\Delta S_{\cal S}=S_{\cal N}(z+\Delta z)-S_{\cal N}(z),
\end{eqnarray}
where $\Delta S_{\cal S}$, from Eq. (\ref{env}), is given by

\begin{eqnarray}
\label{ds2}
\Delta S_{\cal S}= 2\pi k_B \frac{m c}{\hbar}\, \Delta z.
\end{eqnarray}
Eqs. (\ref{enm}) and (\ref{ds2}) show that if we consider initially $\rho_{\cal{N}}(z)$ as a pure state, so we have $S_{\cal{N}}(z)=0$. As Eq. (\ref{ds2}) is different from zero when $\Delta z\neq 0$, then relation (\ref{enm}) represents a mixed state $\rho_{\cal N}(z+\Delta z)$ under the translation along $z$

\begin{eqnarray}
\label{rg}
\rho_{\cal N}(z+\Delta z)\equiv U \rho_{\cal N}(z) U^+.
\end{eqnarray}
Eq. (\ref{rg}) is only satisfied if the translation operator $U\equiv e^{- i \Delta z \hat{P}_z}$ is not an unitary operator, $U U^+ \neq 1$ and consequently, the generator of $z$ translations is not an hermitian operator, $\hat{P}^+_z \neq \hat{P}_z$. At first sight, this is an important result from KA because the time-independent Schr\"odinger equation derived from the translation operator predicts eigenvalues and eigenvectors whose physical results do not agree with the ultra-cold neutron experiments.

It is well known that BG statistical formalism or the standard entanglement entropy property (which is also described by Von Neumann entropy) have strong subadditivity property.  This property, at first sight, turns KA compatible with VE formalism. However, when a physical system presents strong interactions or entanglement, BG formalism, or Von Neumann entropy, can not satisfy an important thermodynamics property that is extensivity\cite{entropy}. Therefore, BG thermostatistics can fail.

There are, at moment, potential entropy candidates to generalize the statistical mechanics of a physical system. The most popular are Renyi\cite{renyi} and Tsallis\cite{Tsa3,jt,abe} theories. Renyi approach, unlike Tsallis, does not satisfy important thermodynamics properties as 
concavity and convexity for $q > 1$\cite{abe}. Therefore, Renyi entropy can not be an appropriate quantity for generalizing BG statistical mechanics. Then, in principle, TT formalism seems to be the unique precise theory for generalizing  BG statistical mechanics.


A system is said to be extensive if

\begin{eqnarray}
\label{ext}
0 < \lim_{N\rightarrow \infty}\, \frac{S(N)}{N} < \infty,
\end{eqnarray}
where $S$ is the entropy and  $N$ is the number of constituents of a system. In order to guarantee that the extensivity property of a system continues to be satisfied when long-range interactions are present, Tsallis has proposed \cite{Tsa,Tsa1,Tsa2,Tsa3} a generalized entropy formula given by

\begin{eqnarray}
\label{nes}
S_q =  k_B \, \frac{1 - \sum_{i=1}^W p_i^q}{q-1}\;\;\;\;\;\; (\sum_{i=1}^W p_i = 1),
\end{eqnarray}
where $p_i$ is the probability of the system of being in a microstate, $W$ is the total number of configurations and $q$ is a real parameter called entropic index. TT formalism contains an additive Boltzmann-Gibbs statistics as a particular case in the limit $ q \rightarrow 1$. The entropy (\ref{nes}) can also be written in terms of the density operator $\rho$ 

\begin{eqnarray}
\label{nes2}
S_q =  k_B \, Tr \frac{\rho - \rho^q}{q-1}.
\end{eqnarray}
From Eq. (\ref{nes}) or (\ref{nes2})  we have an important result for the entropy, namely

\begin{eqnarray}
\label{nae}
S_q(A+B)=S_q(A)+S_q(B)\nonumber \\+\frac{(1-q)}{k_B}S_q(A)S_q(B),
\end{eqnarray}
where we can observe that the entropy is no more additive except for the case $q=1$. TT  nonadditive entropy is applied to a large number of physical systems which are characterized by the existence of long-range correlation.  To mention some examples, there are systems endowed with long-duration memory, anomalous diffusion, turbulence in pure electron plasma and self-graviting systems.  It is worth to mention that, in another study, one of us \cite{JAN} has applied the TT formalism in VE theory specifically in the context of the fundamental length and Newton's law of gravity and to cosmology \cite{aa2}.

\bigskip

In order to circumvent the negative objections pointed out by KA, we will consider that the entropy of neutron's fragments and the remaining bits that constitute the holographic screen, Eq.(\ref{ena}), are no longer additive. Here we are assuming that the quantum neutron fragments state is entangled with the remaining quantum bits state. In this case,  the total entropy of the holographic screen is not a sum of each part. Considering that TT entropy is the most suitable choice for our model, then we can use  Eq. (\ref{nae}) for $q \geq 1$, leading to the relation

\begin{eqnarray}
\label{nbna}
S_{\cal S}(z) \leq S_{\cal S/N}(z) + S_{\cal N}(z).
\end{eqnarray}
Following the same steps of  KA, except condition (\ref{ena}), we obtain

\begin{eqnarray}
\label{dsm}
\Delta S_{\cal S}\geq S_{\cal N}(z+\Delta z) - S_{\cal N}(z).
\end{eqnarray}
We must mention in Eq.(\ref{dsm}) that the equal sign refers to the $\Delta z=0$ case. This equal sign is necessary to maintain (\ref{dsm}) consistent with (\ref{ds2}) for $\Delta\,z=0$.  Consequently, we have that $\Delta S_{\cal S} > 0$ for $\Delta z\neq 0$. If we start with a pure neutron state, i.e., $S_{\cal N}(z)=0$, then the neutron state $\rho_{\cal N}(z+\Delta z) $ can also remain in a pure state and consequently, $S_{\cal N}(z+\Delta z)=0$, a result that does not contradict the inequality $(\ref{dsm})$.  This result makes the translation operator $U\equiv e^{- i \Delta z \hat{P}_z}$ an unitary one and, consequently, the generator of $z$ translations is now an hermitian operator. In other words, $ U^{\dagger} U\,=\,1 \Longrightarrow \hat{P}^+_z  = \hat{P}_z$. The time-independent Schr\"odinger equation now predicts the results that agree nicely with the ultra-cold neutron-gravity experiment as mentioned in KA description. 

\bigskip

To conclude, KA is not the only possible quantum extension of VE formalism. Other possibilities may be interesting options for future studies. 
However, our intention in this work is to analyze  the KA main result in the light of TT ideas together with the quantum mechanical analysis of the holographic screen in order to demonstrate VE hypothesis.
In fact,  bits play an important role in VE formalism. Therefore, it is reasonable that we can generalize the VE theory and assume that the neutron fragments and the remaining bits can interact strongly constituting an entangled state. Consequently, according to TT formalism, the entropy is no longer an additive quantity and the usual Boltzmann-Gibbs thermodynamics is not the most appropriate choice. TT formalism appears as a natural candidate to describe, in a more general way, the thermodynamics of  physical systems that present strong interactions among its constituent. 

\bigskip

EMCA would like to thank Conselho Nacional de Desenvolvimento Cient\' ifico e Tecnol\'ogico (CNPq), a brazilian research support agency, for partial financial support.

\end{document}